\documentclass[prd,a4paper,superscriptaddress,nofootinbib]{revtex4}
\usepackage{amsmath,epsfig,bm}

\begin{document}

%%%%%%%%%%%%%%%%%%%%%%%%%%%%%%%%%%%%%%%%%%%%%%%%%%%%%%%%%%%%%%%%%%
%                       Own commands                             %
%%%%%%%%%%%%%%%%%%%%%%%%%%%%%%%%%%%%%%%%%%%%%%%%%%%%%%%%%%%%%%%%%%

\newcommand{\ve}{\mathbf{e}}
\newcommand{\vnhat}{\hat{\mathbf{n}}}
\newcommand{\vm}{\mathbf{m}}
\newcommand{\vv}{\mathbf{v}}
\newcommand{\vvhat}{\hat{\mathbf{v}}}
\newcommand{\half}{{\textstyle \frac{1}{2}}}
\newcommand{\clp}{{\mathcal{P}}}

%%%%%%%%%%%%%%%%%%%%%%%%%%%%%%%%%%%%%%%%%%%%%%%%%%%%%%%%%%%%%%%%%%
%                       Title etc.                               %
%%%%%%%%%%%%%%%%%%%%%%%%%%%%%%%%%%%%%%%%%%%%%%%%%%%%%%%%%%%%%%%%%%

\title{Peculiar velocity effects in high-resolution microwave background
experiments}

\author{Anthony Challinor}
 \email{A.D.Challinor@mrao.cam.ac.uk}
 \affiliation{Astrophysics Group, Cavendish Laboratory, Madingley Road, Cambridge CB3 OHE, UK.}
\author{Floor {v}an Leeuwen}
 \email{fvl@ast.cam.ac.uk}
 \affiliation{Institute of Astronomy, Madingley Road, Cambridge,
CB3 0HA, UK.}

\begin{abstract}
\vspace{\baselineskip}
We investigate the impact of peculiar velocity effects due to the motion
of the solar system relative to the microwave background (CMB) on high
resolution CMB experiments. It is well known that on the largest angular
scales the combined effects of Doppler shifts and aberration are important;
the lowest Legendre multipoles of total intensity receive power from
the large CMB monopole in transforming from the CMB frame. On small angular
scales aberration dominates and is shown here to lead to significant
distortions of the total intensity and polarization multipoles in transforming
from the rest frame of the CMB to the frame of the solar system.
We provide convenient analytic
results for the distortions as series expansions in the relative velocity of
the two frames, but at the highest resolutions a numerical quadrature is
required. Although many of the high resolution multipoles themselves are
severely distorted by the frame transformations, we show that their statistical
properties distort by only an insignificant amount. Therefore, cosmological
parameter estimation is insensitive to the transformation from the CMB
frame (where theoretical predictions are calculated) to the rest frame of
the experiment.
\end{abstract}

%\pacs{??}

\maketitle

%%%%%%%%%%%%%%%%%%%%%%%%%%%%%%%%%%%%%%%%%%%%%%%%%%%%%%%%%%%%%%%%%%
%                         Main body                              %
%%%%%%%%%%%%%%%%%%%%%%%%%%%%%%%%%%%%%%%%%%%%%%%%%%%%%%%%%%%%%%%%%%

\section{Introduction}
\label{sec:intro}

The impressive advances being made in sensitivity and resolution of microwave
background (CMB) experiments demand that careful attention be paid to
potential systematic effects in the analysis pipeline.
Such effects can arise from imperfect modelling
of the instrument, e.g.\ approximations in modelling the
beam~\cite{challinor00,wu01,souradeep01,fosalba01},
or incomplete knowledge of the pointing, but also from more fundamental
effects such as inaccurate separation of foregrounds (see
e.g.\ Refs.~\cite{bouchet99,hobson99} for reviews).
In this paper we consider errors that may arise due to
neglect of the peculiar motion of the experiment relative to the CMB rest
frame (that frame in which the CMB dipole vanishes). For short duration
experiments (e.g. balloon flights such as MAXIMA~\cite{maxima} and
BOOMERANG~\cite{boomerang}) the relative velocity is
constant over the timescale of the experiment, but for experiments conducted
over a few months or longer, and particularly for satellite
surveys~\cite{cobe,map,planck}, the variation in the relative velocity adds
additional complications. In principle, the modulation of the aberration
arising from any variation in the relative velocity must be accounted for
with a more refined pointing model for the
experiment~\cite{vanleeuwen02,challinor02} when making a map.

For a relative speed of $\beta c$ (where $c$ is the speed of
light and $\beta\sim 1.23\times 10^{-3}$ for the solar-system barycenter
relative to the CMB frame), the r.m.s.\ photon Doppler shifts and deflection
angles are $\beta/\sqrt{3}$ and $\sqrt{2/3}\beta$ respectively. Despite these
small values, significant distortions of the spherical multipoles of the
total intensity and polarization fields do arise. A well known example is
provided by the CMB dipole seen on earth, which, given the observed
spectrum, arises from the transformation of the monopole in the CMB frame.
More generally, on the largest angular scales the combined effects of Doppler
shifts and aberration couple the total intensity monopole and dipole into the
$l$th multipoles at the level $O(\beta^l)$ and $O(\beta^{l-1})$ respectively.
Given the size of the non-cosmological monopole, annual modulation of the
dipole by the variation in the relative velocity of the earth in the CMB
frame must be considered in long duration experiments.

In this paper we
concentrate on the effects of peculiar velocities on small angular scale
features in the microwave sky. On such scales, aberration dominates the
distortions and becomes particularly acute when the angular scales of
interest, $O(1/l)$, drop below the r.m.s.\ deflection angle,
i.e.\ $l \agt 800$ for the transformation from the CMB frame to that
of the solar system. We provide simple analytic results for these distortions
to the total intensity and polarization fields as power series in the relative
velocity $\beta$. The power series converge rather slowly at the highest
multipoles for most values of the azimuthal index $m$ [the
leading-order corrections go like $O(l\beta)$] but the distortions
can still easily be found semi-analytically with a one-dimensional quadrature.
If the transformations of the multipoles carried through to their statistical
properties, theoretical power spectra computed in linear theory (e.g.\ with
standard Boltzmann codes~\cite{seljak96,LCL99}) would not accurately describe
the statistics
of the high resolution multipoles observed on earth. (The theoretical
power spectra would still be accurate in the CMB frame.) It is
straightforward to calculate the statistical correlations of the multipoles
observed on earth. Fortunately, as we show here, the statistical corrections
due to peculiar velocity effects turn out to be negligible despite the large
corrections to the individual multipoles. It follows that for the purposes
of high resolution power spectrum and parameter estimation, the
transformation from the CMB frame can be neglected.

This paper is arranged as follows. In Sec.~\ref{sec:intensity} we describe
the transformation laws for the total intensity multipoles in specific
intensity and frequency-integrated forms. Convenient series expansions in
$\beta$ of the transformations are provided, and their properties under
rotations
of the reference frames are described. The statistical poperties of the
transformed multipoles are investigated by constructing rotationally-invariant
power spectrum estimators and full correlation matrices. In
Sec.~\ref{sec:polarization} we discuss the geometry of the frame
transformations for linear polarization, and present power series expansions
for the transformations of the multipoles. The behaviour under rotations and
parity are also outlined.
Power spectra estimators and correlation matrices are constructed,
and cross correlations with the total intensity are considered. Some
implications of our results for survey missions are discussed in
Sec.~\ref{sec:discussion}, which is followed by our conclusions in
Sec.~\ref{sec:conclusion}. An appendix provides details
of the evaluation of the multipole transformations as power series in
$\beta$.

We use units with $c=1$.

\section{Transformation laws for total intensity}
\label{sec:intensity}

We consider the microwave sky as seen by two observers at the same event.
Observer $S$ is equipped with a comoving tetrad $\{(e_\mu)^a\}$,
$\mu=\{0,1,2,3\}$,
and observer $S'$ carries the Lorentz-boosted tetrad $\{(e_\mu^\prime)^a\}$.
The relative velocity of $S'$ as seen by $S$ has components on $\{(e_i)^a\}$,
$i=\{1,2,3\}$, which we denote by the spatial vector $\vv$, which has
magnitude $\beta$. The $S$ observer receives a photon with four-momentum
$p^a$ when their line of sight is along $\vnhat$, so the photon propagation
direction is $-\vnhat$. For $S$ the photon frequency is $\nu$ where
$h \nu = p_a (e_0)^a$ ($h$ is Planck's constant), while $S'$ observes
frequency
\begin{equation}
\nu' = \nu \gamma (1+\vnhat \cdot \vv),
\label{eq:1}
\end{equation}
where $\gamma^{-2} = 1 - \beta^2$. The line of sight in $S'$ is
\begin{equation}
\vnhat' = \left(\frac{\vnhat \cdot \vvhat + \beta}{1+\vnhat\cdot\vv}\right)
\vvhat + \frac{\vnhat - \vnhat\cdot\vvhat \vvhat}{\gamma(1+\vnhat \cdot \vv)},
\label{eq:2}
\end{equation}
where $\vvhat$ is a unit vector in the direction of the relative velocity.

Denoting the sky brightness in total intensity seen by $S$ as $I(\nu,\vnhat)$,
the brightness seen by $S'$ is (e.g.\ Ref.~\cite{mtw})
\begin{equation}
I'(\nu',\vnhat') = I(\nu,\vnhat) \left(\frac{\nu'}{\nu}\right)^3.
\label{eq:3}
\end{equation}
If $S$ and $S'$ use their spatial triads $\{(e_i)^a\}$ and $\{(e_i^\prime)^a\}$
to define polar coordinates in the usual manner, and expand the sky brightness
in terms of scalar spherical harmonics, i.e.\ $I(\nu,\vnhat)=
\sum_{lm}a_{lm}^I(\nu) Y_{lm}(\vnhat)$, we find the following transformation
law for the brightness multipoles:
\begin{equation}
a^{I\prime}_{lm}(\nu') = \sum_{l'm'} \int \text{d} \vnhat \,
\gamma (1+\vnhat\cdot\vv) a_{l'm'}^I(\nu)
Y_{l'm'}(\vnhat) Y_{lm}^\ast(\vnhat'),
\label{eq:4}
\end{equation}
where $\nu=\nu'\gamma^{-1}(1+\vnhat\cdot \vv)^{-1}$, and
we have used $\nu'{}^2 \text{d}\vnhat' = \nu^2 \text{d}\vnhat$.

It will prove more convenient to consider the integral of the brightness over
frequency, $I(\vnhat) \equiv \int \text{d}\nu\, I(\nu,\vnhat)$. The
transformation law for this flux per solid angle follows from integrating
Eq.~(\ref{eq:3}):
\begin{equation}
I'(\vnhat') = I(\vnhat) \left(\frac{\nu'}{\nu}\right)^4.
\label{eq:5}
\end{equation}
Expanding $I(\vnhat)$ in spherical harmonics, we find the multipole
transformation law
\begin{eqnarray}
a^{I\prime}_{lm} &=& \sum_{l'm'} a^I_{l'm'} \int \text{d} \vnhat \,
[\gamma(1+\vnhat\cdot\vv)]^2 
Y_{l'm'}(\vnhat) Y_{lm}^\ast(\vnhat'), \nonumber \\
&=& \sum_{l'm'} K_{(lm)(lm)'}a^I_{l'm'}
\label{eq:7}
\end{eqnarray}
where $a^I_{lm} = \int \text{d}\nu\, a^I_{lm}(\nu)$. The second equality
defines the kernel $K_{(lm)(lm)'}$ which relates the frequency-integrated
multipoles in $S$ and $S'$. Dividing $a^I_{lm}$ by four times the average
flux per solid angle gives the multipoles of the gauge-invariant temperature
anisotropy in linear theory (e.g.\ Refs.~\cite{stoeger95b,maartens95}).

\begin{figure}
\epsfig{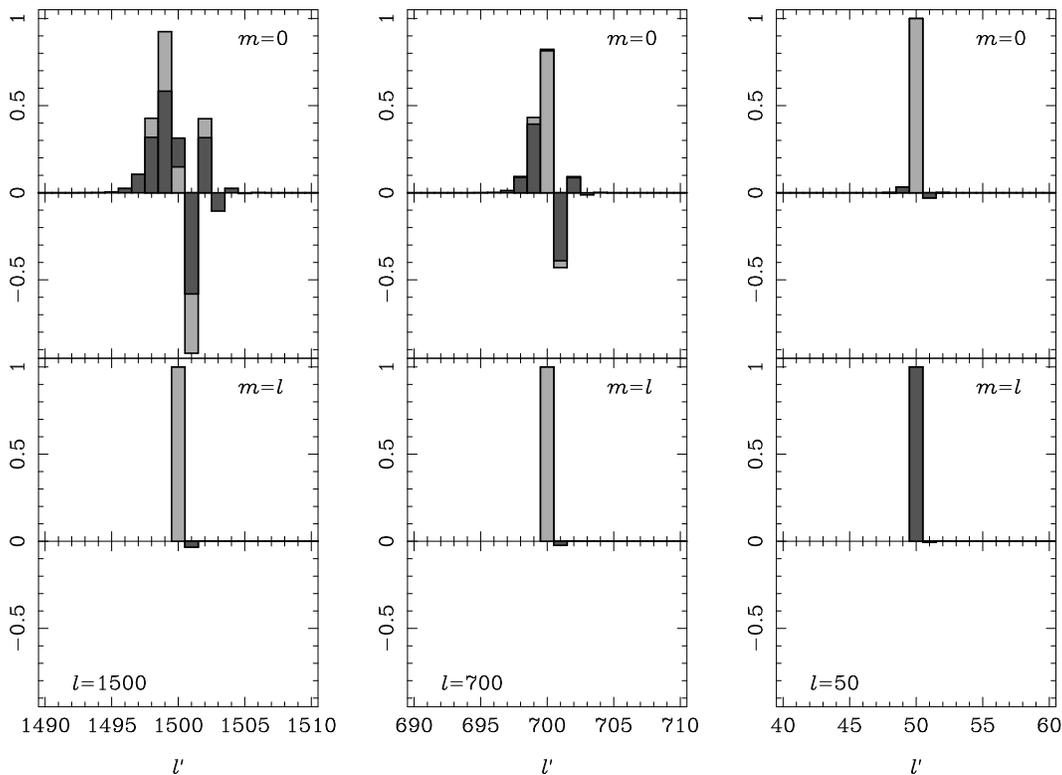}
\caption{\label{fig:one}%
Representative elements of the frequency-integrated kernel
$K_{(lm)(lm)'}$ evaluated with the relative velocity
($\beta=1.23 \times 10^{-3}$) along $(e_3)^a$. The results of a numerical
integration of Eq.~(\ref{eq:7}) are shown in dark gray, while results based
on the series expansion~(\ref{eq:8}) are shown in light gray. The smaller
(absolute values) of the two are shown in the foreground. Elements are shown
for $l=1500$ (left), $l=700$ (middle), and $l=50$ (right), with
$m=0$ (top) and $m=l$ (bottom).}
\end{figure}

If we choose the spacelike vectors of the tetrad $\{(e_\mu)^a\}$ so that the
relative velocity is along $(e_3)^a$, the multipole transformation law becomes
block-diagonal, $K_{(lm)(lm)'} \propto \delta_{mm'}$,
with no coupling between different $m$ modes. The
kernel for a general configuration can then be inferred from
its transformation properties under rotations described in
Sec.~\ref{subsec:rotation}. In the appendix we evaluate Eq.~(\ref{eq:4}) as a
series expansion in $\beta$ for general spin-weight functions,
including terms up to $O(\beta^2)$, for the
case where $\vv$ is aligned with $(e_3)^a$. The expression is cumbersome,
partly due to the fact that the transformation law is non-local in frequency.
For large $l$ the aberration effect dominates Doppler shifts and the
frequency spectrum of the multipoles is preserved by the transformation.
We also give the result obtained by integrating over frequency;
setting $s=0$ in Eq.~(\ref{eq:app7}) we find
the series expansion of the kernel $K_{(lm)(lm)'}$ up to $O(\beta^2)$:
\begin{eqnarray}
K_{(lm)(l'm)} &=&
\delta_{ll'} \biggl[1 + \frac{1}{2}\beta^2 \biggl(C_{(l+1)m}^2(l-1)(l-2)
+C_{lm}^2 (l+2)(l+3) + m^2 - l(l+1) + 2 \biggr) \biggr] \nonumber \\
&&\mbox{}+ \delta_{l(l'+1)} \beta C_{lm}(l+3)
- \delta_{l(l'-1)} \beta C_{(l+1)m}(l-2) \nonumber \\
&&\mbox{} + \delta_{l(l'+2)} \beta^2 C_{lm} C_{(l-1)m} \frac{1}{2}(l+2)(l+3)
+ \delta_{l(l'-2)} \beta^2 C_{(l+2)m} C_{(l+1)m} \frac{1}{2}(l-1)(l-2),
\label{eq:8}
\end{eqnarray}
where $C_{lm} \equiv {}_0 C_{lm}$ with
\begin{equation}
{}_s C_{lm} \equiv \sqrt{\frac{(l^2-m^2)(l^2-s^2)}{l^2(4l^2-1)}}.
\label{eq:9}
\end{equation}
Comparison with Eq.~(\ref{eq:app6}) shows that for high $l$ the aberration
effect described by the term $Y_{lm}^\ast(\vnhat')$ in Eqs.~(\ref{eq:4})
and (\ref{eq:7}) is dominant. For $\beta l \agt 1$ the series is slow to
converge for $|m| \ll l$ since the leading-order corrections go like
$O(l \beta)$, reflecting the fact that the deflection angle due to aberration
is comparable to the angular scale of the spherical harmonics at this $l$.
For $\beta \approx 1.23\times 10^{-3}$, appropriate for the
solar-system barycenter relative to the CMB frame, $\beta l \sim 1$
corresponds to multipoles $l \sim 800$. In this case, the kernel
$K_{(lm)(lm)'}$ is easily evaluated by a numerical quadrature. We show
some representative elements of the kernel in Fig.~\ref{fig:one}, which
demonstrates that the multipoles do indeed suffer severe distortion for
$l \agt 1/\beta$, as suggested by Eq.~(\ref{eq:8}).
For $|m|$ close to $l$, ${}_s
C_{lm} \sim O(l^{-1/2})$ and retaining only the terms given
in Eq.~(\ref{eq:8}) is accurate to much better than 0.1 per cent for the
$l$ range probed by e.g.\ Planck ($l \alt 2000$). For such values of $m$
the distortions to the multipoles are only small, with leading-order
corrections at $l'=l\pm 1$ of $O(\beta \sqrt{l})$. For $l \beta \ll 1$,
the departures of the kernel from the identity $\delta_{ll'}\delta_{mm'}$
are very small, giving negligible distortions to the multipoles except for
$l$ close to unity when the non-zero coupling to the (large) monopole can give
significant distortions, as described in Sec.~\ref{sec:intro}.

%%%%%%%%%%%%%%%%%%%%%%%%%%%%%%%%%%%%%%%%%%%%%%%%%%%%%%%%%%%%
%%%%%%%%%%%%%%%%%%%%%%%%%%%%%%%%%%%%%%%%%%%%%%%%%%%%%%%%%%%%

\subsection{Rotational properties}
\label{subsec:rotation}

If we rotate the relative velocity $\vv$ to $D \vv$\footnote{Here,
$D$ denotes the appropriate representation of the rotation group $SO(3)$.}
keeping the tetrad $(e_\mu)^a$ fixed,
[thus inducing a transformation of the Lorentz-boosted tetrad
$(e_\mu^\prime)^a$], the frequency-integrated multipoles continue to be
given by Eq.~(\ref{eq:7}), but with $\vv$ replaced by $D\vv$ in $\vnhat'$
[Eq.~(\ref{eq:2})] and in $(1+\vnhat \cdot \vv)$. With the change of
integration variable $\vnhat \rightarrow D \vnhat$, the integral defining
the transformed kernel $K_{(lm)(lm)'}(D \vv)$ becomes
\begin{equation}
\int \text{d}\vnhat \, \gamma^2 (1+\vnhat\cdot \vv)Y_{l'm'}(D \vnhat)
Y_{lm}^\ast(D \vnhat') = \int \text{d}\vnhat \, \gamma^2 (1+\vnhat\cdot \vv)
D^{-1} Y_{l'm'}(\vnhat) [D^{-1} Y_{lm}(\vnhat')]^\ast,
\label{eq:10}
\end{equation}
where $D^{-1} Y_{l'm'}(\vnhat) = \sum_{m''} D^{l\ast}_{m'm''}
Y_{lm''}(\vnhat)$ with $D^l_{mm'}$ a Wigner $D$-matrix. (Our conventions
for $D$-matrices follow Refs.~\cite{brink93,varshalovich88}.)
It follows that the transformed kernel is given by
\begin{equation}
K_{(lm)(lm)'}(D \vv) = \sum_{MM'}
D^l_{m M}{} K_{(lM)(lM)'}(\vv) D^{l'\ast}_{m'M'}.
\label{eq:11}
\end{equation}

Instead of rotating the (physical) relative velocity of $S$ and $S'$, we could
imagine rotating the spatial triad $(e_i)^a \rightarrow D (e_i)^a$. Under this
coordinate transformation, the Lorentz-boosted frame vectors transform
similarly: $(e'_i)^a \rightarrow D (e'_i)^a$. For a fixed sky, the
multipoles seen by $S$ and $S'$ transform according to e.g.\
$a^I_{lm} \rightarrow \sum_{m'}D^{l\ast}_{m'm} a^I_{lm'}$
(which is equivalent to rotating the sky with $D^{-1}$ leaving the tetrad
fixed). It follows that under coordinate rotations, the kernel transforms
as
\begin{equation}
K_{(lm)(lm)'} \rightarrow D^{l\ast}_{Mm} K_{(lM)(lM)'} D^{l'}_{M'm'}.
\label{eq:12}
\end{equation}
Note that the (passive) rotation of the frame vectors by $D^{-1}$ has the
same effect on the kernel as the (active) rotation of the
relative velocity $\vv$ by $D$, as expected.

Finally, we consider (active) parity transformations $\vv\rightarrow-\vv$ with
the tetrad $(e_\mu)^a$ held fixed. Using $Y_{lm}(-\vnhat) = (-1)^l
Y_{lm}(\vnhat)$ it is straightforward to show that
\begin{equation}
K_{(lm)(lm)'}(-\vv) = (-1)^{l+l'} K_{(lm)(lm)'}(\vv).
\label{eq:13}
\end{equation}
The behaviour of the kernel under parity ensures that if we simultaneously
invert $\vv$ and the sky [$a^I_{lm} \rightarrow (-1)^l a^I_{lm}$], the
multipoles seen by $S'$ transform to $(-1)^l a^{I\prime}_{lm}$.

These transformation properties of the kernel under rotations allow one
to generalise Eq.~(\ref{eq:8}) easily to the case where $\vv$ is not aligned
with $(e_3)^a$.

\subsection{Power spectrum estimators}
\label{subsec:power}

We have seen how aberration effects lead to significant distortions
of some of the high-$l$ multipoles in transforming from the CMB frame to the
frame of the experiment. In the next two subsections we investigate the
impact of these distortions on the statistical properties of the multipoles.

We assume that in the CMB frame ($S$) the second-order statistics of the
anisotropies are summarised by
\begin{equation}
\langle a^I_{lm} a^{I\ast}_{l'm'} \rangle =
C^{II}_l \delta_{ll'} \delta_{mm'},
\label{eq:14}
\end{equation}
appropriate to a statistically-isotropic ensemble with power spectrum
$C^{II}_l$. (The averaging is over an ensemble of CMB realizations.)
It is this $C_l$ for $l\geq 2$ that is computed with linear perturbation
theory in standard Boltzmann codes (e.g.\ Refs.~\cite{seljak96,LCL99}).

We begin by considering the quadratic statistic
\begin{equation}
\hat{C}^{II\prime}_l \equiv \frac{1}{2l+1}\sum_{m} |a^{I\prime}_{lm}|^2,
\label{eq:15}
\end{equation}
which is evaluated by $S'$. In the absence of noise this statistic is the
optimal (minimum-variance) estimator for the power spectrum if we ignore
peculiar velocity effects. By construction, $\hat{C}^{II\prime}_l$
is independent of
the choice of spatial triad, but is only invariant under rotations of the sky
in $S$ ($a^I_{lm} \rightarrow D^l_{mm'} a_{lm'}^I$) if the relative velocity
is also rotated to $D\vv$. However, averaging over CMB realizations
\emph{keeping the relative velocity fixed} we obtain a quantity
$\langle \hat{C}^{II\prime}_l \rangle$ which is
obviously invariant under
rotations of the sky in $S$. The average $\langle \hat{C}^{II\prime}_l
\rangle$, which determines the
bias of the power spectrum estimator $\hat{C}^{II\prime}_l$, is
linearly related to the $C^{II}_l$:
\begin{eqnarray}
\langle \hat{C}^{II\prime}_l \rangle &=& \frac{1}{2l+1} \sum_{l' m m'}
|K_{(lm)(lm)'}|^2 C^{II}_{l'} \nonumber \\
           &\equiv& \sum_{l'} W_{ll'} C^{II}_{l'}.
\label{eq:16}
\end{eqnarray}
The kernel $W_{ll'}$ depends only on the relative speed $\beta$ and not the
direction $\vvhat$, so we can always evaluate it with  $\vv$ aligned with
$(e_3)^a$.

%%%%%%%%%%%%%%%%%%%%%%%%%%%%%%%%%%%%%%%%%%%%%%%%%%%%%%%%%%%%
\begin{figure}
\epsfig{figure=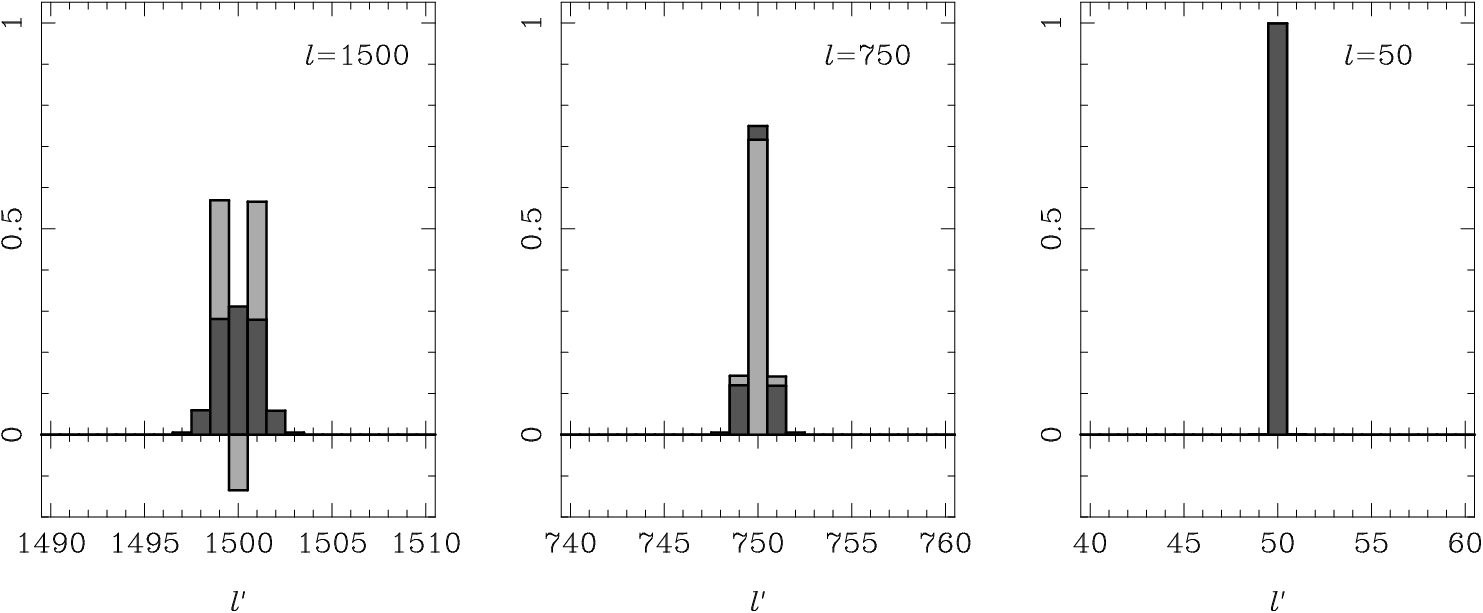,angle=-90,width=14cm}
\caption{\label{fig:two}%
Representative elements of the kernel
$W_{ll'}$ evaluated with relative velocity
$\beta=1.23 \times 10^{-3}$. The results of a numerical
integration of are shown in dark gray, while results based
on the series expansion~(\ref{eq:17}) are shown in light gray. The smaller
(absolute values) of the two are shown in the foreground. Elements are shown
for $l=1500$ (left), $l=700$ (middle), and $l=50$ (right).}
\end{figure}
%%%%%%%%%%%%%%%%%%%%%%%%%%%%%%%%%%%%%%%%%%%%%%%%%%%%%%%%%%%%

The series expansion~(\ref{eq:8}) of $K_{(lm)(lm')}$ can be used to evaluate
$W_{ll'}$. Correct to $O(\beta^2)$ we find
\begin{equation}
W_{ll'} = \delta_{ll'} \left(1- \frac{1}{3}\beta^2(l^2+l-8)\right)
+ \delta_{l(l'+1)}\beta^2 \frac{l(l+3)^2}{3(2l+1)} 
+ \delta_{l(l'-1)} \beta^2\frac{(l-2)^2(l+1)}{3(2l+1)}.
\label{eq:17}
\end{equation}
Again the series is slow to converge for $l\beta \agt 1$ and the terms
neglected in Eq.~(\ref{eq:17}) are non-negligible.
We show $W_{ll'}$ for some representative $l$ values in Fig.~\ref{fig:two}.
It is clear from the figure that $W_{ll'}$ is well localised in comparison to
any features in the CMB power spectrum for the range of $\beta$ of interest
here. In this case, we can approximate
\begin{eqnarray}
\langle\hat{C}^{II'}_l\rangle &\approx& C^{II}_l \sum_{l'} W_{ll'} \nonumber \\
          &=& C^{II}_l [1+4\beta^2+O(\beta^3)].
\label{eq:18}
\end{eqnarray}
The effect of the velocity transformation is thus to rescale the amplitude
of the power spectrum by $1+4\beta^2$. This bias is clearly insignificant.
$\sum_{l'}W_{ll'}$ is actually independent of $l$ to all
orders in $\beta$. To see this we form $\sum_{l'}W_{ll'}$ directly using the
integral expression (\ref{eq:7}) for $K_{(lm)(lm)'}$. The result
simplifies to
\begin{equation}
\sum_{l'}W_{ll'}= \frac{\gamma^4}{4\pi}\int \text{d}\vnhat\, (1+\vnhat\cdot
\vv)^4 = \gamma^4 \left(1+2\beta^2 + \frac{1}{5}\beta^4\right),
\label{eq:19}
\end{equation}
on using the completeness relation
\begin{equation}
\sum_{lm} {}_s Y_{lm}(\vnhat_1) {}_s Y_{lm}^\ast(\vnhat_2)
= \delta(\vnhat_1-\vnhat_2),
\label{eq:20}
\end{equation}
and the addition theorem for the spherical harmonics. The series expansion
of Eq.~(\ref{eq:19}) agrees with Eq.~(\ref{eq:18}). We conclude that despite
the fact that the multipoles themselves can be severely distorted by
aberration for $l\beta\agt 1$ in passing from the CMB frame to that of the solar
system, the quadratic power spectrum estimator is negligibly biased since the
effect of the velocity transformation is to convolve the power spectrum with
a narrow kernel $W_{ll'}$ that sums to very nearly unity.

\subsection{Signal covariance matrix}
\label{subsec:covariance}

Assuming Gaussian statistics in the CMB frame, the multipoles
$a^{I\prime}_{lm}$ in $S'$ will also be distributed according to
a multivariate Gaussian since the transformation~(\ref{eq:7}) is linear.
In this case, the covariance matrix $\langle a^{I\prime}_{lm}
a^{I\prime\ast}_{l'm'} \rangle$ contains all statistical information
about the anisotropies in $S'$, and as such is an essential element of
optimal power spectrum estimation.

If we make use of Eq.~(\ref{eq:14}), the covariance matrix in $S'$ reduces
to
\begin{equation}
\langle a^{I\prime}_{lm}a^{I\prime\ast}_{l'm'} \rangle =
\sum_{LM} K_{(lm)(LM)}K^\ast_{(lm)'(LM)} C_{L}^{II}.
\label{eq:21}
\end{equation}
The presence of the preferred direction $\vvhat$ breaks statistical
isotropy in $S'$, and the multipoles are correlated for $l\neq l'$ and
$m\neq m'$. The structure of the covariance matrix in $S'$ depends on
the choice of the spatial triad $(e_i)^a$ with respect to the relative
velocity of the two observers. Aligning $(e_3)^a$ with $\vv$, the $m$-modes
decouple in $K_{(lm)(lm)'}$ and so also in the covariance matrix. Furthermore,
for the values of $\beta$ of interest here ($\beta \ll 1$), the kernel
$K_{(lm)(lm)'}$ falls rapidly to zero for $l$ and $l'$ differing by more
than a few (see Fig.~\ref{fig:one}), so the same will be true of the
covariance matrix. It follows that we can approximate
\begin{equation}
\langle a^{I\prime}_{lm}a^{I\prime\ast}_{l'm'} \rangle \approx
C^{II}_l \sum_{LM} K_{(lm)(LM)}K^\ast_{(lm)'(LM)}.
\label{eq:22}
\end{equation}
(Pulling out $C^{II}_{l'}$ instead will give essentially the same result for
smooth power spectra.) The summation in Eq.~(\ref{eq:22}) is most easily
evaluated by substituting the integral representation~(\ref{eq:7}) for
$K_{(lm)(LM)}$ and using the completeness relation~(\ref{eq:20}).
We find that $\sum_{LM} K_{(lm)(LM)}K^\ast_{(lm)'(LM)}$ reduces to the
integral
\begin{equation}
\int \text{d}\vnhat\, [\gamma(1+\vnhat\cdot \vv)]^4 Y_{lm}^\ast(\vnhat')
Y_{l'm'}(\vnhat') = \int \text{d}\vnhat'\, [\gamma(1-\vnhat'\cdot\vv)]^{-6}
Y_{lm}^\ast(\vnhat') Y_{l'm'}(\vnhat'),
\label{eq:23}
\end{equation}
where we changed the integration variable to $\vnhat'$ and used
$\gamma(1+\vnhat\cdot \vv) = [\gamma(1-\vnhat'\cdot\vv)]^{-1}$. Note that
both spherical harmonics have the same argument in the integrand, so we
don't expect the same $O(l\beta)$ terms at high $l$ that arise in
the kernel $K_{(lm)(lm)'}$. Eq.~(\ref{eq:23}) can easily be evaluated for
$(e_3)^a$ along $\vv$ (in which case there is no coupling between different
$m$) by expanding in $\beta$:
\begin{eqnarray}
\sum_{LM} K_{(lm)(LM)}K^\ast_{(l'm)(LM)} &=& \delta_{ll'}[1 +
3\beta^2(7 C_{(l+1)m}^2 + 7C_{lm}^2 - 1)] + \delta_{l(l'+1)}
6\beta C_{lm} + \delta_{l(l'-1)} 6 \beta C_{(l+1)m} \nonumber \\
&&\mbox{} + \delta_{l(l'+2)} 21 \beta^2 C_{lm}C_{(l-1)m} + \delta_{l(l'-2)}
21 \beta^2 C_{(l+2)m}C_{(l+1)m} + O(\beta^3). 
\label{eq:24}
\end{eqnarray}
This result for the covariance matrix in $S'$ could easily be used in
maximum-likelihood power spectrum estimation (see e.g.\ Ref.~\cite{bond98}) to
correct for the bias due to peculiar velocity effects. However,
since the leading corrections are only $O(\beta)$, even at high $l$, the
effects will be negligible.

\section{Transformation laws for linear polarization}
\label{sec:polarization}

The linearly polarized brightness in $S$ is described by Stokes parameters
$Q(\nu,\vnhat)$ and $U(\nu,\vnhat)$. The Stokes parameters depend on
a specific choice of orthonormal basis vectors $\{\vm_1,\vm_2\}$ for each line
of sight $\vnhat$. If $\{\vm_1,\vm_2,-\vnhat\}$ form a right-handed
orthonormal set, the Stokes parameters are related to the linear
polarization tensor by
\begin{equation}
\clp^{ab} = \frac{1}{2}[Q(\vm_1 \otimes \vm_1 - \vm_2 \otimes \vm_2)
+ U(\vm_1 \otimes \vm_2 + \vm_2 \otimes \vm_1)].
\label{eq:25}
\end{equation}
The Stokes parameters transform under changes of frame in the same way as the
total intensity, i.e.\
\begin{equation}
Q'(\nu',\vnhat') = Q(\nu,\vnhat) \left(\frac{\nu'}{\nu}\right)^3,
\label{eq:26}
\end{equation}
and similarly for $U$, provided that the basis vectors are transformed
according to~\cite{chall99c}
\begin{equation}
\vm'_i =  \vm_i + (\gamma-1) \vm_i\cdot \vvhat \vvhat 
- \gamma  \vm_i \cdot \vv \vnhat',
\label{eq:27}
\end{equation}
where $i=1,2$.
It is straightforward to verify that this transformation law preserves
orthonormality, and also that $\vm'_i$ is obtained from $\vm_i$ by parallel
transport on the unit sphere along the great circle through $\vnhat$ and
$\vnhat'$ (and so through $\vvhat$ also). In terms of the polarization
tensor, the frame transformation law can be written as
\begin{equation}
\clp^{\prime ab} (\nu',\vnhat') = \clp^{ab}_\| (\nu,\vnhat;\vv)
\left(\frac{\nu'}{\nu}\right)^3,
\label{eq:28}
\end{equation}
where $\clp^{ab}_\| (\nu,\vnhat;\vv)$ is $\clp^{ab} (\nu,\vnhat)$
parallel propagated to $\vnhat'$. The 1+3 covariant form of this transformation
was given in Ref.~\cite{chall99c}.

If $S$ and $S'$ introduce polar coordinates as in Sec.~\ref{sec:intensity},
the polarization tensor can be expanded in symmetric trace-free tensor
harmonics~\cite{kamion97c}:\footnote{Our $a^E_{lm}$ and $a^B_{lm}$ are
$\sqrt{2}$ times the gradient ($G$) and curl ($C$) multipoles introduced in
Refs.~\cite{kamion97c,kamion97}. With this convention the power spectra of the
electric and magnetic multipoles agree with those defined
in the spin-weight formalism~\cite{seljak97,zaldarriaga97}.}
\begin{equation}
\clp_{ab}(\nu,\vnhat) = \frac{1}{\sqrt{2}}\sum_{lm}
a^E_{lm}(\nu)Y^G_{(lm)ab}(\vnhat) + a^B_{lm}(\nu) Y^C_{(lm)ab}(\vnhat),
\label{eq:29}
\end{equation}
which defines the electric ($E$) and magnetic ($B$) multipoles. Using
Eq.~(\ref{eq:28}) we can extract the multipoles seen by $S'$. For
$a^{E\prime}_{lm}(\nu')$ we find
\begin{equation}
a^{E\prime}_{lm}(\nu') = \sum_{l'm'} \int\text{d}\vnhat\,\gamma(1+\vnhat\cdot
\vv) [a^E_{l'm'}(\nu') Y^{G\, ab}_{\| (lm)'}(\vnhat;\vv)
Y^{G\ast}_{(lm)ab}(\vnhat') + a^B_{l'm'}(\nu')
Y^{C\, ab}_{\| (lm)'}(\vnhat;\vv) Y^{G\ast}_{(lm)ab}(\vnhat')],
\label{eq:30}
\end{equation}
with a similar result for $a^{B\prime}_{lm}(\nu')$.
Here $Y^{G\, ab}_{\| (lm)}(\vnhat;\vv)$ is $Y^{G\, ab}_{(lm)}(\vnhat)$
parallel propagated to $\vnhat'$, and similarly for the curl harmonics.
Note how in general the frame transformation mixes $E$ and $B$
polarization. Equation~(\ref{eq:30}) is valid quite generally, and is useful
for discussing the rotational properties of the transformations (see later).
However, to compute the transformation laws it is again convenient
to arrange $(e_i)^a$ so that $\vv$ is along $(e_3)^a$. We can then
exploit the fact that the polar basis vector fields
$\hat{\bm{\theta}}(\vnhat)$ and
$\hat{\bm{\phi}}(\vnhat)$ are parallel propagated along longitudes to
simplify $Y^{G\, ab}_{\| (lm)}(\vnhat;\vv)$. The gradient and curl harmonics
can be written in terms of spin-weight $\pm 2$ harmonics
(our conventions follow Refs.~\cite{ng99,challinor00}):
\begin{eqnarray}
Y^{G\, ab}_{lm} &=& \frac{1}{\sqrt{2}}({}_{-2}Y_{lm} \vm\otimes\vm
+ {}_2 Y_{lm} \vm^\ast \otimes \vm^\ast), \label{eq:31} \\
Y^{C\, ab}_{lm} &=& \frac{1}{i\sqrt{2}}({}_{-2}Y_{lm} \vm\otimes\vm
- {}_2 Y_{lm} \vm^\ast \otimes \vm^\ast), \label{eq:32}
\end{eqnarray}
where the complex vector $\vm \equiv (\hat{\bm{\theta}}+i\hat{\bm{\phi}})
/\sqrt{2}$, so that Eq.~(\ref{eq:30}) can be written as
\begin{equation}
(a^{E\prime}_{lm} \pm i a^{B\prime}_{lm})(\nu') = \sum_{l'm'}
(a^E_{l'm'} \pm i a^B_{l'm'})(\nu)
\int \text{d}\vnhat \,\gamma(1+\vnhat\cdot \vv) 
{}_{\pm 2}Y_{l'm'}(\vnhat){}_{\pm 2}Y_{lm}^\ast(\vnhat').
\label{eq:33}
\end{equation}
The integral on the right-hand side is evaluated as a power series in
$\beta$ for general spin-weight $s$ in the appendix.

For our purposes it will be more convenient to consider the
frequency-integrated multipoles, e.g.\
$a^E_{lm} = \int \text{d}\nu\, a^E_{lm}(\nu)$. Integrating Eq.~(\ref{eq:30})
over frequency, we find
\begin{eqnarray}
a^{E\prime}_{lm} &=& \sum_{l'm'}{}_+ K_{(lm)(lm)'} a^E_{l'm'}
+ i {}_- K_{(lm)(lm)'} a^B_{l'm'}, \label{eq:34}\\
a^{B\prime}_{lm} &=& \sum_{l'm'}{}_+ K_{(lm)(lm)'} a^B_{l'm'}
- i {}_- K_{(lm)(lm)'}a^E_{l'm'}, \label{eq:35}
\end{eqnarray}
where the kernels
\begin{eqnarray}
{}_+ K_{(lm)(lm)'} &=& \int \text{d}\vnhat\, [\gamma(1+\vnhat\cdot\vv)]^2
Y^{G\, ab}_{\| (lm)'}(\vnhat;\vv)Y^{G\ast}_{(lm)ab}(\vnhat')
= \int \text{d}\vnhat\, [\gamma(1+\vnhat\cdot\vv)]^2
Y^{C\, ab}_{\| (lm)'}(\vnhat;\vv)Y^{C\ast}_{(lm)ab}(\vnhat'),
\label{eq:36}\\
{}_- K_{(lm)(lm)'} &=& -i\int \text{d}\vnhat\, [\gamma(1+\vnhat\cdot\vv)]^2
Y^{C\, ab}_{\| (lm)'}(\vnhat;\vv)Y^{G\ast}_{(lm)ab}(\vnhat')
= i\int \text{d}\vnhat\, [\gamma(1+\vnhat\cdot\vv)]^2
Y^{G\, ab}_{\| (lm)'}(\vnhat;\vv)Y^{C\ast}_{(lm)ab}(\vnhat') .
\label{eq:37}
\end{eqnarray}

The behaviour of ${}_\pm K_{(lm)(lm)'}$ under rotations $\vv \rightarrow
D\vv$ is the same as for the total intensity kernel, Eq.~(\ref{eq:11}),
since the tensor harmonics transform under rigid rotations with the same
$D$-matrices as the scalar harmonics~\cite{challinor00}.
This property of the tensor
harmonics also ensures that under rotations of the coordinate system,
$(e_i)^a \rightarrow D (e_i)^a$, the electric and magnetic multipoles
transform irreducibly to e.g.\ $\sum_{m'} D^{l\ast}_{m'm}{} a^E_{lm'}$.
Under inversion of $\vv$ with $(e_\mu)^a$ held fixed, the kernels
transform to
\begin{eqnarray}
{}_+ K_{(lm)(lm)'}(-\vv) &=& (-1)^{l+l'} {}_+ K_{(lm)(lm)'}(\vv),
\label{eq:38}\\
{}_- K_{(lm)(lm)'}(-\vv) &=& -(-1)^{l+l'} {}_- K_{(lm)(lm)'}(\vv),
\label{eq:39}
\end{eqnarray}
so that under simultaneous inversion of the sky in $S$,
$a^E_{lm} \rightarrow (-1)^l a^E_{lm}$ and $a^B_{lm} \rightarrow
(-1)^{l+1} a^B_{lm}$, and inversion of $\vv$, the multipoles in
$S'$ transform like those in $S$.

The frequency-integrated kernels are most simply evaluated with $\vv$ along
$(e_3)^a$. In this case the $m$-modes decouple, as with the total intensity.
Writing ${}_\pm K = ({}_2 K \pm {}_{-2} K)/2$, we can use Eq.~(\ref{eq:app7}),
which evaluates ${}_s K_{(lm)(lm)'}$ as a series correct to $O(\beta^2)$, to
show that
\begin{eqnarray}
{}_+ K_{(lm)(l'm)} &=& \delta_{ll'}\biggl[1+\frac{1}{2}\beta^2\biggl(
{}_2 C_{(l+1)m}^2(l-1)(l-2)
+{}_2 C_{lm}^2 (l+2)(l+3) + m^2 - l(l+1) +6
-\frac{4m^2}{l(l+1)}+\frac{24m^2}{l^2(l+1)^2}\biggr)\biggr]\nonumber\\
&+& \delta_{l(l'+1)} \beta {}_2 C_{lm} (l+3)
- \delta_{l(l'-1)} \beta {}_2 C_{(l+1)m}(l-2)\nonumber \\
&+& \delta_{l(l'+2)}\beta^2 {}_2 C_{lm} {}_2 C_{(l-1)m} \frac{1}{2}
(l+2)(l+3)
+ \delta_{l(l'-2)}\beta^2 {}_2 C_{(l+2)m} {}_2 C_{(l+1)m} \frac{1}{2}
(l-1)(l-2), \label{eq:40}
\end{eqnarray}
and
\begin{equation}
{}_- K_{(lm)(l'm)} = - \delta_{ll'} \frac{6\beta m}{l(l+1)}
- \delta_{l(l'+1)} {}_2 C_{lm} (l+3)\frac{6\beta^2 m}{(l+1)(l-1)}
+ \delta_{l(l'-1)} {}_2 C_{(l+1)m}(l-2)\frac{6\beta^2 m}{l(l+2)}.
\label{eq:41}
\end{equation}
(Equivalent results, correct to $O(\beta)$, have already been worked out
in 1+3 covariant form~\cite{chall99c}.)
The kernel ${}_- K_{(lm)(lm)'}$ is suppressed at
high $l$. It receives comparable contributions from Doppler and aberration
effects for all $l$ [see Eq.~(\ref{eq:app6})] in contrast to the
${}_+ K_{(lm)(lm)'}$ and the total intensity kernel
$K_{(lm)(lm)'}$ which are dominated by aberration effects at high $l$.
The series expansion of ${}_+ K_{(lm)(lm)'}$ is slow to converge for
$l \beta \agt 1$ when $|m| \ll l$, and there are large distortions to the
electric and magnetic multipoles for these indices. Electric multipoles
nearby in $l$ couple in strongly to distort $a^E_{lm}$, and similarly
for the magnetic multipoles. For $l \gg 1$ the kernel ${}_+K_{(lm)(lm)'}$
is almost indistinguishable from the total intensity kernel $K_{(lm)(lm)'}$.
The cross
contamination of e.g.\ $B$ by $E$ due to the frame transformation is much
weaker, with the maximal effect $\sim O(\beta/l)$ at leading order occuring for
$|m| \approx l$. [Note that, as  with ${}_+ K_{(lm)(lm)'}$, the convergence of
of Eq.~(\ref{eq:41}) is slow for $l\beta \agt 1$ when $|m| \ll l$.]
The transfer of power from $E$ to $B$ is potentially the
most interesting effect since in the absence of astrophysical foregrounds,
inflationary models predict that magnetic polarization in the CMB frame
on scales larger than a degree or so arises only from gravitational waves.
However, on these scales a gravity wave background comprising only one percent
of the large-angle temperature anisotropy would have $B$ power far in excess
of that generated in the frame of the
experiment by transforming $E$ from the CMB frame. On sub-degrees scales,
where any primordial $B$-polarization is expected to be very small,
other non-linear effects, most
notably weak lensing of $E$~\cite{zaldarriaga98b}, will dominate the
$B$ signal produced by the velocity transformation.

\subsection{Power spectrum estimators}
\label{subsec:power_pol}

The second-order statistics of the polarization multipoles in the CMB frame,
assuming statistical isotropy and parity invariance, define power spectra:
\begin{eqnarray}
\langle a^E_{lm} a^{E\ast}_{l'm'} \rangle &=& \delta_{ll'}\delta_{mm'}
C_l^{EE}, \label{eq:42}\\
\langle a^B_{lm} a^{B\ast}_{l'm'} \rangle &=& \delta_{ll'}\delta_{mm'}
C_l^{BB}, \label{eq:43}\\
\langle a^E_{lm} a^{I\ast}_{l'm'} \rangle &=& \delta_{ll'}\delta_{mm'}
C_l^{IE}, \label{eq:44}
\end{eqnarray}
with no correlations between $B$ and $E$ or $I$. We can form
estimators of these power spectra from the multipoles
in $S'$ by analogy with Eq.~(\ref{eq:15}), e.g.\
\begin{equation}
\hat{C}^{IE\prime}_l = \frac{1}{2l+1}\sum_{m} a^{E\prime}_{lm}
a^{I\prime\ast}_{lm}.
\label{eq:45}
\end{equation}
Since these estimators are rotationally invariant we can compute them
for $\vv$ aligned with $(e_3)^a$ using Eqs.~(\ref{eq:40}) and (\ref{eq:41})
without loss of generality.

\begin{figure}
\epsfig{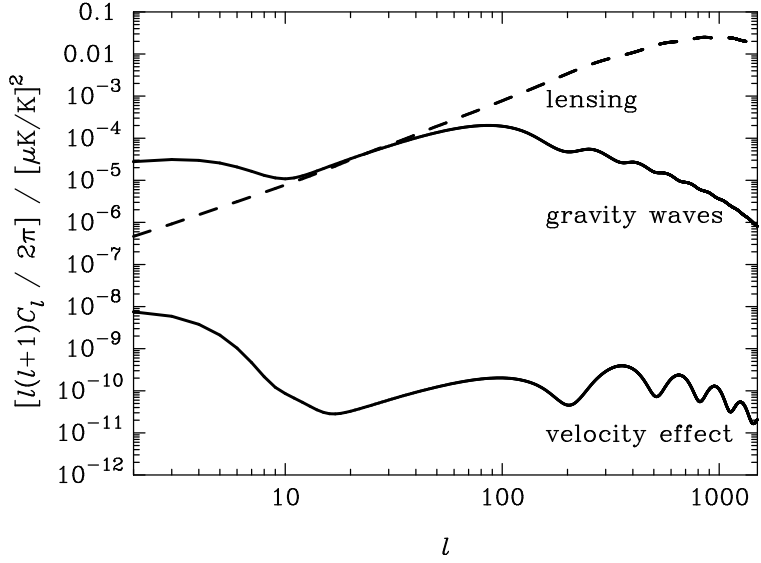}
\caption{\label{fig:three}%
Contribution of $C_l^{EE}$ to the mean estimator $\langle
\hat{C}_l^{BB\prime} \rangle$ in a $\Lambda$CDM model with
one percent contribution to the total-intensity quadrupole from gravity
waves. This velocity effect is compared with $C_l^{BB}$ (in the CMB frame)
due to primordial gravity waves (solid line) and weak lensing of the
$E$-polarization (dashed line).}
\end{figure}

The expected values of the power spectra estimators can be expressed in terms
of the power spectra in the CMB frame using Eqs.~(\ref{eq:7}), (\ref{eq:34}),
and (\ref{eq:35}):
\begin{eqnarray}
\langle \hat{C}^{EE\prime}_l \rangle &=& \frac{1}{2l+1}\sum_{l'm'm}
|{}_+ K_{(lm)(lm)'}|^2 C^{EE}_{l'}
+ |{}_- K_{(lm)(lm)'}|^2 C^{BB}_{l'}, \label{eq:46} \\
\langle \hat{C}^{BB\prime}_l \rangle &=& \frac{1}{2l+1}\sum_{l'm'm}
|{}_+ K_{(lm)(lm)'}|^2 C^{BB}_{l'}
+ |{}_- K_{(lm)(lm)'}|^2 C^{EE}_{l'}, \label{eq:47} \\
\langle \hat{C}^{IE\prime}_l \rangle &=& \frac{1}{2l+1}\sum_{l'm'm}
{}_+ K_{(lm)(lm)'} K_{(lm)(lm)'}^\ast C^{IE}_{l'}. \label{eq:48}
\end{eqnarray}
Substituting the power series expressions for the kernels and performing
the summations over $m$ and $m'$ we find
\begin{eqnarray}
\frac{1}{2l+1}\sum_{mm'}|{}_+ K_{(lm)(lm)'}|^2 &=&
\delta_{ll'}\biggl(1-\frac{1}{3}\beta^2(l+4)(l-3)\biggr)
+ \delta_{l(l'+1)} \beta^2 \frac{(l+3)^2(l^2-4)}{3l(2l+1)} \nonumber\\
&&\mbox{} + \delta_{l(l'-1)} \beta^2 \frac{(l-2)^2(l+3)(l-1)}{3(l+1)(2l+1)},
\label{eq:49} \\
\frac{1}{2l+1}\sum_{mm'}|{}_- K_{(lm)(lm)'}|^2 &=&
\delta_{ll'} \beta^2 \frac{12}{l(l+1)}, \label{eq:50}
\end{eqnarray}
and
\begin{eqnarray}
\frac{1}{2l+1}\sum_{mm'} {}_+ K_{(lm)(lm)'} K_{(lm)(lm)'}^\ast
&=& \delta_{ll'}\biggl(1-\frac{1}{3}\beta^2(l^2+l-10)\biggr)
+ \delta_{l(l'+1)} \beta^2\sqrt{l^2-4}\frac{(l+3)^2}{3(2l+1)} \nonumber\\
&&\mbox{}+ \delta_{l(l'-1)} \beta^2\sqrt{(l+3)(l-1)}\frac{(l-2)^2}{3(2l+1)},
\label{eq:51}
\end{eqnarray}
correct to $O(\beta^2)$. For $l \gg 1$ the right-hand sides of
Eqs.~(\ref{eq:49}) and (\ref{eq:51}) are almost equal to each other and to
the kernel $W_{ll'}$ which determines the bias in the total-intensity
estimator $\hat{C}_l^{II'}$. As with the total intensity, the series
in Eqs.~(\ref{eq:49}--\ref{eq:51}) are slow to converge for $l \beta \agt 1$.
The bias of $\hat{C}_l^{BB\prime}$ by $E$-polarization is controlled by
$\sum_{mm'} |{}_- K_{(lm)(lm)'}|^2 / (2l+1)$, which falls off rapidly with
$l$. In Fig.~\ref{fig:three} we compare this contribution to the expected
$\langle \hat{C}_l^{BB\prime} \rangle$ with the $B$-polarization power spectrum
due to primordial gravity waves and weak lensing of the $E$-polarization.
The cosmological model is a Lambda, cold dark matter ($\Lambda$CDM) in which
gravity waves contribute one percent to the large-angle temperature anisotropy.
As remarked earlier, the contamination arising from the frame transformation
is well below the expected $C_l^{BB}$ in such a model.

%%%%%%%%%%%%%%%%%%%%%%%%%%%%%%%%%%%%%%%%%%%%%%%%%%%%%%%%%%%%
%%%%%%%%%%%%%%%%%%%%%%%%%%%%%%%%%%%%%%%%%%%%%%%%%%%%%%%%%%%%

The means of the estimators
$\hat{C}^{EB\prime}_l$ and $\hat{C}^{EB\prime}_l$, defined by analogy with
$\hat{C}^{IE\prime}_l$, would vanish in the absence of peculiar velocity
effects (and foregrounds) due to parity. The velocity transformations
preserve these zero means since
\begin{equation}
\sum_{mm'} {}_+ K_{(lm)(lm)'} {}_- K^\ast_{(lm)(lm)'}
= \sum_{mm'} {}_- K_{(lm)(lm)'} K^\ast_{(lm)(lm)'} = 0. 
\label{eq:52}
\end{equation}
These results are easily proved by choosing $\vv$ along $(e_3)^a$ so that all
kernels are real, and using the general results
${}_\pm K_{(lm)(lm)'}^\ast = \pm (-1)^{m+m'} {}_\pm K_{(l\, -m)(l'\, -m')}$
and $K_{(lm)(lm)'}^\ast = (-1)^{m+m'} K_{(l\, -m)(l'\, -m')}$.

The kernels represented by the left-hand sides of
Eqs.~(\ref{eq:49})--(\ref{eq:51}) fall off sufficiently rapidly with
$|l'-l|$ that they are narrow compared to expected features in the primordial
power spectra~\footnote{At low $l$ the polarization power spectra vary rapidly
(as power laws) with $l$. Over this part of the spectrum the approximation
that the power is approximately constant over the width of the convolving
kernel is still valid since the latter are essentially Kronecker deltas
at low $l$.}. Following the analysis in Sec.~\ref{subsec:power} we can pull
out $C_{l'}^{EE}$, $C_{l'}^{BB}$, and $C_{l'}^{IE}$ at $l'=l$ from
the summations in Eqs.~(\ref{eq:46})--(\ref{eq:48}). Performing the sums over
$l'$, we find
\begin{eqnarray}
\frac{1}{2l+1}\sum_{mm'l'}|{}_+ K_{(lm)(lm)'}|^2 &=&
1 + 4 \beta^2 \frac{l^2+l-3}{l(l+1)},
\label{eq:53} \\
\frac{1}{2l+1}\sum_{mm'l'}|{}_- K_{(lm)(lm)'}|^2 &=& \beta^2 \frac{12}{l(l+1)}
, \label{eq:54} \\
\frac{1}{2l+1}\sum_{mm'l'} {}_+ K_{(lm)(lm)'} K_{(lm)(lm)'}^\ast
&=& 1 + \frac{1}{3}\beta^2\left[\sqrt{(l+3)(l-1)}\frac{(l-2)^2}{2l+1}
+ \sqrt{l^2-4}\frac{(l+3)^2}{2l+1} - l^2-l+10\right],
\label{eq:55}
\end{eqnarray}
correct to $O(\beta^2)$. For large $l$ the right-hand sides of
Eqs.~(\ref{eq:53}) and (\ref{eq:55}) approach $1+4 \beta^2$; as with the
total intensity, there is a negligible scaling of the amplitude of
the power spectra estimated in the $S'$ frame due to the frame transformation.
Note that
\begin{eqnarray}
\frac{1}{2l+1}\sum_{mm'l'} [|{}_+ K_{(lm)(lm)'}|^2 +|{}_- K_{(lm)(lm)'}|^2]
&=& \frac{1}{2(2l+1)}\sum_{mm'l'} [|{}_2 K_{(lm)(lm)'}|^2 +|{}_{-2}
K_{(lm)(lm)'}|^2]  \nonumber \\
&=& \gamma^4 \left(1 + 2\beta^2 + \frac{1}{5}\beta^4\right), \label{eq:56}
\end{eqnarray}
where we have used the completeness relation and addition
theorem  for the spin-$s$ harmonics. Adding Eqs.~(\ref{eq:53}) and
(\ref{eq:54}) we obtain the series expansion of the exact result in
Eq.~(\ref{eq:56}).

\subsection{Signal covariance matrices}
\label{subsec:covariance_pol}

The calculation of the covariance matrix of the polarization multipoles
in $S'$ follows that for the total intensity given in
Sec.~\ref{subsec:covariance}. For smooth power spectra we can approximate
\begin{eqnarray}
\langle a^{E\prime}_{lm} a^{E\prime\ast}_{l'm'} \rangle &\approx&
C_l^{EE} \sum_{LM} {}_+ K_{(lm)(LM)} {}_+ K^\ast_{(lm)'(LM)}
+ C_l^{BB} \sum_{LM} {}_- K_{(lm)(LM)} {}_- K^\ast_{(lm)'(LM)} ,
\label{eq:57}\\
\langle a^{B\prime}_{lm} a^{B\prime\ast}_{l'm'} \rangle &\approx&
C_l^{BB} \sum_{LM} {}_+ K_{(lm)(LM)} {}_+ K^\ast_{(lm)'(LM)}
+ C_l^{EE} \sum_{LM} {}_- K_{(lm)(LM)} {}_- K^\ast_{(lm)'(LM)}
\label{eq:58},\\
\langle a^{E\prime}_{lm} a^{I\prime\ast}_{l'm'} \rangle &\approx&
C_l^{IE} \sum_{LM} {}_+ K_{(lm)(LM)} K^\ast_{(lm)'(LM)}.
\label{eq:59}
\end{eqnarray}
The remaining correlators would vanish for $\vv=0$ due to parity invariance.
For non-zero $\vv$ we can approximate
\begin{eqnarray}
\langle a^{E\prime}_{lm} a^{B\prime\ast}_{l'm'} \rangle &\approx&
i C_l^{EE} \sum_{LM} {}_+ K_{(lm)(LM)} {}_- K^\ast_{(lm)'(LM)}
+ i C_l^{BB} \sum_{LM} {}_- K_{(lm)(LM)} {}_+ K^\ast_{(lm)'(LM)},
\label{eq:60} \\
\langle a^{B\prime}_{lm} a^{I\prime\ast}_{l'm'} \rangle &\approx&
-i C_l^{IE} \sum_{LM} {}_- K_{(lm)(LM)} K^\ast_{(lm)'(LM)}.
\label{eq:61}
\end{eqnarray}
If we align $\vv$ with $(e_3)^a$ we can evaluate these expressions by
substituting for the series expansions of the kernels from Eqs.~(\ref{eq:40})
and (\ref{eq:41}). The $m$ modes decouple and we find
\begin{eqnarray}
\sum_{LM}{}_+ K_{(lm)(LM)} {}_+ K^\ast_{(l'm)(LM)} &=&
\delta_{ll'} \left[1 + 3 \beta^2\left(7 {}_2 C_{(l+1)m}^2 + 7 {}_2
C_{lm}^2 + \frac{16m^2}{l^2(l+1)^2} - 1\right)\right] \nonumber \\
&&\mbox{} + \delta_{l(l'+1)}
6 \beta {}_2 C_{lm} \nonumber + \delta_{l(l'-1)} 6 \beta {}_2 C_{(l+1)m}
\nonumber \\
&&\mbox{}+ \delta_{l(l'+2)} 21 \beta^2 {}_2C_{lm}{}_2C_{(l-1)m}
+ \delta_{l(l'-2)} 21\beta^2 {}_2 C_{(l+2)m} {}_2 C_{(l+1)m}, \label{eq:62} \\
\sum_{LM}{}_- K_{(lm)(LM)} {}_- K^\ast_{(l'm)(LM)} &=&
\delta_{ll'} \frac{36 \beta^2 m^2}{l^2(l+1)^2}, \label{eq:63}
\end{eqnarray}
and
\begin{eqnarray}
\sum_{LM}{}_+ K_{(lm)(LM)} K^\ast_{(l'm)(LM)} &=&
\delta_{ll'}\biggl(1+ \frac{1}{2}\beta^2[-(l+1)(l+19)({}_2 C_{(l+1)m}^2
+ C_{(l+1)m}^2) - l(l-18)({}_2 C_{lm}^2 + C_{lm}^2) \nonumber \\
&&\mbox{} + 18 + 2(l-2)^2 {}_2 C_{(l+1)m} C_{(l+1)m}
+ 2 (l+3)^2 {}_2 C_{lm} C_{lm}]\biggr) \nonumber \\ 
&&\mbox{} - \delta_{l(l'+1)} \beta[(l-3) C_{lm} - (l+3){}_2 C_{lm}]
+ \delta_{l(l'-1)} \beta[(l+4) C_{(l+1)m} - (l-2) {}_2 C_{(l+1)m}] \nonumber \\
&&\mbox{} + \frac{1}{2}\delta_{l(l'+2)}\beta^2[(l+2)(l+3){}_2 C_{lm}
{}_2 C_{(l-1)m}
+ (l-3)(l-4) C_{lm} C_{(l-1)m} \nonumber \\
&&\mbox{} - 2 (l+3)(l-4) {}_2 C_{(l-1)m} C_{lm}]
+ \frac{1}{2}\delta_{l(l'-2)} \beta^2[(l-1)(l-2){}_2 C_{(l+2)m}{}_2 C_{(l+1)m}
\nonumber \\
&&\mbox{}
+ (l+4)(l+5) C_{(l+2)m}C_{(l+1)m} - 2(l-2)(l+5) {}_2 C_{(l+1)m} C_{(l+2)m}],
\label{eq:64}
\end{eqnarray}
correct to $O(\beta^2)$. This final expression is cumbersome and hides the
fact that the leading order corrections to the covariance matrices are
only $O(\beta)$, rather than $O(\beta l)$. To see this, we can expand
Eq.~(\ref{eq:64}) in $1/l$ for large $l$ to find
\begin{eqnarray}
\sum_{LM}{}_+ K_{(lm)(LM)} K^\ast_{(l'm)(LM)} &=&
\delta_{ll'} \left[1+ \beta^2\left(\frac{15}{2} - \frac{155 + 84 m^2}{8l^2}
\right)\right] + \delta_{l(l'+1)}\beta \left(3 - \frac{1}{l} -
\frac{3(7 + 4m^2)}{8l^2} \right) \nonumber \\
&&\mbox{} + \delta_{l(l'-1)} \beta
\left(3 + \frac{1}{l} - \frac{29+12m^2}{8l^2}\right)
+ \delta_{l(l'+2)} \beta^2 \left(\frac{21}{4} - \frac{5}{2l} -
\frac{159+84 m^2}{16 l^2} \right) \nonumber \\
&&\mbox{} + \delta_{l(l'-2)} \beta^2
\left(\frac{21}{4} + \frac{7}{2l} - \frac{223 + 84m^2}{16 l^2}\right),
\label{eq:65}
\end{eqnarray}
correct to $O(l^{-2})$. For $|m| \approx l$ the expansion in $1/l$ is slow to
converge, and the full expression, Eq.~(\ref{eq:64}), should be evaluated
exactly if the (very small) corrections to the covariance matrices are to be
included in a statistical analysis. It is worth noting that
\begin{equation}
\begin{split}
\sum_{LM}({}_+ K_{(lm)(LM)} {}_+ K^\ast_{(l'm)(LM)}
&+ {}_- K_{(lm)(LM)} {}_- K^\ast_{(l'm)(LM)}) \\ &= 
\frac{1}{2} \int\text{d}\vnhat[\gamma(1-\vnhat\cdot \vv)]^{-6}[
{}_2 Y_{lm}^\ast(\vnhat) {}_2 Y_{l'm}(\vnhat) +
{}_{-2} Y_{lm}^\ast(\vnhat) {}_{-2} Y_{l'm}(\vnhat)],
\end{split}
\label{eq:66}
\end{equation}
for $\vv$ along $(e_3)^a$, where we have used the completeness relation,
Eq.~(\ref{eq:20}). It is straightforward to show with an expansion in $\beta$
that Eq.~(\ref{eq:66}) is consistent with adding Eqs.~(\ref{eq:62}) and
(\ref{eq:63}).

For the correlators $\langle a^{E\prime}_{lm} a^{B\prime\ast}_{l'm'} 
\rangle$ and $\langle a^{B\prime}_{lm} a^{I\prime\ast}_{l'm'} \rangle$,
which would vanish for $\vv=0$, we require the results (for
$\vv$ aligned with $(e_3)^a$)
\begin{eqnarray}
\sum_{LM} {}_+ K_{(lm)(LM)} {}_- K^\ast_{(l'm)(LM)}
&=& - \delta_{ll'} \frac{6 \beta m}{l(l+1)} - \delta_{l(l'+1)}{}_2 C_{lm}
(7l+3) \frac{6\beta^2 m}{l(l-1)(l+1)} \nonumber \\
&&\mbox{} - \delta_{l(l'-1)}{}_2 C_{(l+1)m} (7l+4)
\frac{6 \beta^2 m}{l(l+1)(l+2)}, \label{eq:67}\\
\sum_{LM} {}_- K_{(lm)(LM)}  K^\ast_{(l'm)(LM)}
&=& - \delta_{ll'} \frac{6\beta m}{l(l+1)} - \delta_{l(l'+1)}
[l(l+3) {}_2 C_{lm}-(l-1)(l-3)C_{lm}] \frac{6 \beta^2 m}{l(l+1)(l-1)}
\nonumber \\
&&\mbox{} + \delta_{l(l'-1)} [(l+1)(l-2) {}_2 C_{(l+1)m}
- (l+4)(l+2) C_{(l+1)m}] \frac{6 \beta^2 m}{l(l+1)(l+2)}, \label{eq:68}
\end{eqnarray}
correct to $O(\beta^2)$, and the general result
\begin{equation}
\sum_{LM} {}_- K_{(lm)(LM)} {}_+ K^\ast_{(lm)'(LM)}
= - (-1)^{m+m'} \sum_{LM} {}_+ K_{(l'\, -m')(LM)}
{}_- K^\ast_{(l\, -m)(LM)}.
\label{eq:69}
\end{equation}
The leading order corrections to the components of the correlation matrices
that vanish for $\vv=0$ are $O(\beta)$, and are suppressed at large $l$.
and so can safely be ignored. For completeness we note that
\begin{equation}
\begin{split}
\sum_{LM}({}_+ K_{(lm)(LM)} {}_- K^\ast_{(l'm)(LM)}
&+ {}_- K_{(lm)(LM)} {}_+ K^\ast_{(l'm)(LM)}) \\ &= 
\frac{1}{2} \int\text{d}\vnhat[\gamma(1-\vnhat\cdot \vv)]^{-6}[
{}_2 Y_{lm}^\ast(\vnhat) {}_2 Y_{l'm}(\vnhat) -
{}_{-2} Y_{lm}^\ast(\vnhat) {}_{-2} Y_{l'm}(\vnhat)].
\end{split}
\label{eq:70}
\end{equation}
This result is easily shown to be consistent with Eqs.~(\ref{eq:67}) and
(\ref{eq:69}).

\section{Implications for survey missions}
\label{sec:discussion}

For experiments which observe for less than a month or so the velocity of the
instrument relative to the CMB frame can reasonably be considered constant.
In this case a map in the frame of the instrument can be made with no
account of the effects considered in this paper. Accounting for the
peculiar velocity relative to the CMB frame can be deferred until the
statistical properties of the map are considered. As we have shown here,
peculiar velocity effects can safely be ignored when estimating smooth power
spectra since the estimated power spectra are essentially convolutions of
the spectra in the CMB frame (which we can reliably compute with linear
perturbation theory) with narrow kernels that integrates to unity.

For survey experiments that observe for the order of a year or more the
variation in the orbital velocity of the instrument adds another potential
complication. Modulation of the dipole by the orbital velocity of the earth
was visible in the COBE DMR data~\cite{smoot91};
here we are interested in effects at
small angular scales. To estimate the importance of the effect we consider a
toy model of the Planck High Frequency Instrument (HFI). We approximate the
orbit of the satellite relative to the sun as a linear motion with
$\beta=10^{-4}$ for six months, after which the direction of motion is
reversed for the next six months of observation. Clearly, this toy model will
over estimate the effects of the variation in orbital velocity. Planck will
cover the full sky in six months, so for each six month period we could
make a map and extract the spherical multipoles. In our toy model these
two maps are produced in frames with a relative velocity of
$2\beta = 2\times 10^{-4}$. In the $l$-range relevant to Planck we need only
retain the $O(\beta)$ corrections in Eq.~(\ref{eq:8}), so the difference
between the multipoles measured from the two maps can be approximated as
\begin{equation}
\Delta a^I_{lm} \approx \beta l \sqrt{1-m^2/l^2} (a_{(l-1)m}^I - a_{(l+1)m}^I)
\label{eq:71}
\end{equation}
for large $l$. Here, $a_{lm}^I$ are the total intensity multipoles in the
rest frame of the solar system. The r.m.s.\ difference in the multipoles
is
\begin{equation}
\langle |\Delta a_{lm}^I|^2 \rangle^{1/2} \approx \sqrt{2}
\beta l \sqrt{1-m^2 / l^2} \sqrt{C_l^{II}} \leq \sqrt{2} \beta l
\sqrt{C_l^{II}},
\label{eq:72}
\end{equation}
which should be compared to the instrument noise. For the 100 GHz Planck
HFI channel, the one-year pixel noise is $6.0\, \mu\text{K}$ in 9.2 arcmin
(the beam full-width at half maximum) pixels.
The noise on our six month maps will be larger than this figure by
$\sqrt{2}$. A comparison of the noise on the recovered multipoles with the
r.m.s.\ error due to the difference in orbital velocity shows that the
latter is just above the noise in the region of the first acoustic peak in
$C_l^{II}$ (at $l \sim 200$) for $|m|$ small compared to $l$.
Combining maps at different frequency would
reduce the noise while preserving the peculiar velocity effect. However,
since we have certainly over estimated the importance of the variation in
orbital velocity, it is likely that the variation in aberration due to
the orbital motion of the earth need not be considered beyond the dipole
(which is modulated by the large CMB monopole). In principle, the
modulation of the high $l$ multipoles could easily be accounted for during
map-making by including the aberration corrections in the pointing model of
the instrument~\cite{vanleeuwen02,challinor02}.

\section{Conclusion}
\label{sec:conclusion}

We have shown that for total intensity the
effect of the frame transformation from the CMB frame to that of the solar
system produces large distortions in certain multipoles at high $l$. These
effects arise principally from aberration rather than Doppler
shifts. The linear polarization multipoles are similarly distorted at
high $l$, but with the additional complication that
there is some transfer of power between $E$ and $B$ polarization. This transfer
is suppressed at large $l$, and receives comparable contributions from
aberration and Doppler shifts on all scales.
Although the power in $B$ polarization is expected
to be much smaller than that in $E$ in the absence of foregrounds, the
$B$ polarization generated from $E$ is well below the primordial level
even if gravity waves contribute only one percent of the large-angle
temperature anisotropies. If the gravity wave background is much below this
level, weak gravitational lensing will dominate the primordial signal on all
scales. This lensing signal is expected to be an order of magnitude larger
than the $B$ polarization generated from the frame transformation on large
scales.

Despite significant $O(\beta l)$ distortions of certain multipoles at
large $l$, peculiar velocity effects are suppressed in power spectrum
estimators and the covariance matrices for the CMB signals.
The effect of the frame
transformation on the mean of the simplest power spectrum estimator is to
convolve the spectrum in the CMB frame (which we can compute reliably with
linear perturbation theory) with a narrow kernel that integrates to unity.
For smooth spectra there is negligible bias introduced by such a convolution.
For linear polarization, the bias of e.g.\ the $B$-polarization power spectrum
by $E$ is suppressed at large $l$, and is expected to be negligible on all
scales. We also showed that the frame transformation has only a negligible
effect [$O(\beta)$ as opposed to $O(\beta l)$] on the signal covariance
matrices for smooth underlying power spectra. The leading order effect is
a coupling to the adjacent $l$ values, $l\pm 1$. For linear polarization
additional correlations are induced between $B$ and $E$ polarization,
and $B$ and total intensity $I$, since the frame transformation does not
preserve parity invariance, but their level is negligible.

If the CMB fluctuations are Gaussian in the CMB frame, the multipoles
will remain Gaussian distributed in any other frame since the transformation
is linear in the signal. The transformation does break rotational and parity
invariance, however, and so the aberration effects described here may be
important when searching for weak lensing effects in the microwave
background (using small patches of the sky over a coherence area of the weak
shear), or the effects of non-trivial topologies.

\begin{acknowledgments}
AC thanks the Theoretical Astrophysics Group at Caltech for hospitality while
some of this work was completed, and Marc Kamionkowski for useful
discussions.
AC acknowledges a PPARC Postdoctoral Fellowship; FvL is supported by PPARC.
\end{acknowledgments}

\appendix
\section{Series expansion of the multipole transformation laws}
\label{app:series}

In this appendix we outline the evaluation of the transformation law for the
brightness multipoles as a power series in $\beta$. We align the 
relative velocity with the vector $(e_3)^a$ so that there is no coupling
between different $m$ modes. To allow us to discuss
both total intensity and linear polarization, we consider the integral
\begin{equation}
a^{\prime}_{lm}(\nu') = \sum_{l'} \int \text{d} \vnhat \,
\frac{\nu'}{\nu} a_{l'm}(\nu)
{}_sY_{l'm}(\vnhat) {}_sY_{lm}^\ast(\vnhat'),
\label{eq:app1}
\end{equation}
where $\nu'/\nu = \gamma (1+\vnhat\cdot \vv)$ and $\vnhat'$ is given by
Eq.~(\ref{eq:2}). We Taylor expand $a_{l'm}(\nu)$ as
\begin{equation}
a_{l'm}(\nu) = a_{l'm}(\nu') - \nu'\frac{\text{d}}{\text{d} \nu'}
a_{l'm}(\nu') \left(\beta\mu - \beta^2 \mu^2 + \frac{1}{2}\beta^2\right)
+\frac{\nu'{}^2}{2} \frac{\text{d}^2}{\text{d} \nu'{}^2} a_{l'm}(\nu')
\beta^2 \mu^2 + O(\beta^3),
\label{eq:app2}
\end{equation}
where $\mu \equiv \vnhat \cdot \vvhat$, and we handle ${}_sY_{lm}(\vnhat')$
with the expansion
\begin{equation}
{}_s Y_{lm}(\vnhat') = {}_s Y_{l'm}(\vnhat) -\beta(1-\beta\mu)
(\mu^2-1)\frac{\text{d}}{\text{d}\mu}{}_s Y_{l'm}(\vnhat) 
+\frac{\beta^2 (\mu^2-1)^2}{2} \frac{\text{d}^2}{\text{d}\mu^2}
{}_s Y_{l'm}(\vnhat) + O(\beta^3).
\label{eq:app3}
\end{equation}
The derivatives with respect to $\mu$ can be eliminated with repeated use of
the identity~\cite{varshalovich88}
\begin{equation}
(\mu^2-1)\frac{\text{d}}{\text{d} \mu} {}_s Y_{lm} = l {}_s C_{(l+1)m}
{}_s Y_{(l+1)m} + \frac{sm}{l(l+1)} {}_s Y_{lm}-(l+1){}_s C_{lm}{}_s
Y_{(l-1)m}, \label{eq:app4}
\end{equation}
where ${}_s C_{lm}$ is defined in Eq.~(\ref{eq:9}), and residual
factors of $\mu$ can be absorbed with the identity~\cite{varshalovich88}
\begin{equation}
\mu {}_s Y_{lm} = {}_s C_{(l+1)m} {}_s Y_{(l+1)m} - \frac{sm}{l(l+1)}
{}_s Y_{lm} + {}_s C_{lm} {}_s Y_{(l-1)m}.
\label{eq:app5}
\end{equation}
With these results, we find the following expression for $a'_{lm}(\nu')$:
\begin{eqnarray}
a'_{lm}(\nu') &=& \biggl\{1- \frac{\beta sm}{l(l+1)}\left(
2-\nu'\frac{\text{d}}{\text{d}\nu'}\right) + \beta^2 
\biggl[ \frac{1}{2}{}_s C_{(l+1)m}^2\left(l(l-1)+ 2l
\nu' \frac{\text{d}}{\text{d}\nu'} + \nu'{}^2
\frac{\text{d}^2}{\text{d}\nu'{}^2}\right) \nonumber \\
&&\mbox{} + \frac{1}{2}{}_s C_{lm}^2\left((l+1)(l+2)- 2(l+1) \nu'
\frac{\text{d}}{\text{d}\nu'} + \nu'{}^2
\frac{\text{d}^2}{\text{d}\nu'{}^2}\right)
+ \frac{1}{2}\left(m^2+s^2-l(l+1) + 1 - \nu'\frac{\text{d}}{\text{d}\nu'}
\right) \nonumber \\
&&\mbox{} - \frac{s^2 m^2}{2l(l+1)} + \frac{s^2 m^2}{l^2(l+1)^2}
\left(1 - \nu' \frac{\text{d}}{\text{d}\nu'} + \frac{\nu'{}^2}{2}
\frac{\text{d}^2}{\text{d}\nu'{}^2}\right)\biggr]\biggr\} a_{lm}(\nu')
\nonumber \\
&&\mbox{}-\beta {}_s C_{(l+1)m}\biggl[(l-1) + \nu'\frac{\text{d}}{\text{d}\nu'}
- \frac{\beta sm}{l(l+2)}\left(2(l-1)-(l-2)\nu'\frac{\text{d}}{\text{d}\nu'}
- \nu'{}^2 \frac{\text{d}^2}{\text{d}\nu'{}^2}\right)\biggr] a_{(l+1)m}(\nu')
\nonumber \\
&&\mbox{} - \beta {}_s C_{lm}\biggl[-(l+2) +\nu'\frac{\text{d}}{\text{d}\nu'} 
- \frac{\beta sm}{(l+1)(l-1)}\left(-2(l+2) + (l+3)
\nu'\frac{\text{d}}{\text{d}\nu'} - \nu'{}^2
\frac{\text{d}^2}{\text{d}\nu'{}^2}\right)\biggr] a_{(l-1)m}(\nu') \nonumber \\
&&\mbox{} + \frac{1}{2}\beta^2 {}_s C_{(l+2)m} {}_s C_{(l+1)m} \left(
l(l-1)+2l\nu'\frac{\text{d}}{\text{d}\nu'} + \nu'{}^2
\frac{\text{d}^2}{\text{d}\nu'{}^2}\right) a_{(l+2)m}(\nu') \nonumber \\
&&\mbox{} + \frac{1}{2}\beta^2 {}_s C_{lm} {}_s C_{(l-1)m} \left(
(l+1)(l+2)-2(l+1)\nu'\frac{\text{d}}{\text{d}\nu'} + \nu'{}^2
\frac{\text{d}^2}{\text{d}\nu'{}^2}\right) a_{(l-2)m}(\nu') + O(\beta^3).
\label{eq:app6}
\end{eqnarray}
Integrating this result with respect to $\nu'$, the kernel ${}_s K_{(lm)(lm)'}$
introduced in Sec.~\ref{sec:intensity} ($s=0$) and Sec.~\ref{sec:polarization}
($s=\pm 2$) evaluates to
\begin{eqnarray}
{}_s K_{(lm)(l'm)} &=& \delta_{ll'} \biggl[1 - \frac{3\beta sm}{l(l+1)}
+ \frac{1}{2}\beta^2 \biggl({}_s C_{(l+1)m}^2(l-1)(l-2)+{}_s C_{lm}^2
(l+2)(l+3) \nonumber \\
&&\mbox{} + m^2 + s^2 - l(l+1) + 2 - \frac{s^2m^2}{l(l+1)} + 
\frac{6s^2m^2}{l^2(l+1)^2} \biggr) \biggr] \nonumber \\
&&\mbox{} + \delta_{l(l'+1)} \beta {}_s C_{lm}(l+3)
\left(1-\frac{3\beta sm}{(l+1)(l-1)} \right)
- \delta_{l(l'-1)} \beta {}_s C_{(l+1)m}(l-2)\left(1-
\frac{3\beta sm}{l(l+2)}\right) \nonumber \\
&&\mbox{}+ \delta_{l(l'+2)} \beta^2 {}_s C_{lm} {}_s C_{(l-1)m} \frac{1}{2}
(l+2)(l+3) + \delta_{l(l'-2)} \beta^2 {}_s C_{(l+2)m} {}_s C_{(l+1)m}
\frac{1}{2}(l-1)(l-2) + O(\beta^3),
\label{eq:app7}
\end{eqnarray}
with $K_{(lm)(lm)'}=0$ for $m\neq m'$ in the configuration with $\vv$ along
$(e_3)^a$.

\vspace{\baselineskip}

%%%%%%%%%%%%%%%%%%%%%%%%%%%%%%%%%%%%%%%%%%%%%%%%%%%%%%%%%%%%%%%%%%
%                        References                              %
%%%%%%%%%%%%%%%%%%%%%%%%%%%%%%%%%%%%%%%%%%%%%%%%%%%%%%%%%%%%%%%%%%

\end{document}